%% file: proceed_GayDucati.tex
\documentclass[10pt]{article}
\usepackage{graphicx}

\def\Title#1{\begin{center} {\Large #1 } \end{center}}
\def\Author#1{\begin{center}{ \sc #1} \end{center}}
\def\Address#1{\begin{center}{ \it #1} \end{center}}

\newcommand\pubblock{\rightline{\begin{tabular}{l} Proceedings of the Second Annual LHCP\\ \pubnumber\\
         \pubdate  \end{tabular}}}

\newenvironment{Abstract}{\begin{quotation} \begin{center} 
             \large ABSTRACT \end{center}\bigskip 
      \begin{center}\begin{large}}{\end{large}\end{center} \end{quotation}}

\newenvironment{Presented}{\begin{quotation} \begin{center} 
             PRESENTED AT\end{center}\bigskip 
      \begin{center}\begin{large}}{\end{large}\end{center} \end{quotation}}

\input econfmacros.tex

\usepackage{color}

 \newcommand\pubnumber{ }

\newcommand\pubdate{\today}

\def\affiliation{
Department of Physics \\
Federal University of Rio Grande do Sul, Porto Alegre, Brazil }

\begin{document}

\large
\begin{titlepage}
\pubblock

\vfill
\Title{  DIFFRACTIVE PHOTOPRODUCTION OF $\psi(2S)$ IN PHOTON-POMERON REACTIONS IN PbPb COLLISIONS AT THE CERN LHC  }
\vfill

\Author{ MARIA BEATRIZ GAY DUCATI, MIRIAN THUROW GRIEP AND MAGNO VAL\'ERIO TRINDADE MACHADO}
\Address{\affiliation}
\vfill
\begin{Abstract}

This work investigates the exclusive photoproduction of $\psi(2S)$ meson off nuclei
evaluating the coherent and the incoherent contributions.
The theoretical framework used in the present analysis is the light-cone dipole formalism and
predictions are done for PbPb collisions at the CERN-LHC energy of 2.76 TeV. 
A comparison is done to the recent ALICE Collaboration data
for the $\psi(1S)$ state photoproduction.

\end{Abstract}
\vfill

\begin{Presented}
The Second Annual Conference\\
 on Large Hadron Collider Physics \\
Columbia University, New York, U.S.A \\ 
June 2-7, 2014
\end{Presented}
\vfill
\end{titlepage}
\def\thefootnote{\fnsymbol{footnote}}
\setcounter{footnote}{0}
%

\normalsize 


\section{Introduction}

\hspace{0.5cm}The vector meson photoproduction is sensitive to
the environment and yields new information about the
dynamics of hard processes.
This process provides crucial constraints on properties of the QCD Pomeron 
and the vector meson wave functions and is expected to probe the nuclear gluon-distribution.
The sufficiently large mass of quarkonium states gives the perturbative scale.
 The scattering process is characterized by the color dipole cross section representing the
 interaction of those color dipoles with the target (protons or nuclei). 
The dipole cross section evolution at small Bjorken-$x$
is given by the solution of a non-linear evolution equation.
Dipole sizes of magnitude
 $r\sim 1/\sqrt{m_V^2+Q^2}$ ($m_V$ is the vector meson mass) are probed by the $1S$ vector meson
 production amplitude \cite{nik}.
The production amplitude of radially excited vector mesons $2S$ is suppressed compared with the $1S$
state due to the node effect \cite{Nemchik:1996pp}. 
Which means a strong cancellation of the dipole size contributions
 to the production amplitude from the region above, and below the node position in the $2S$ 
radial wavefunction \cite{Nemchik:2000de}. 
The ratio $\sigma (\psi^{\prime})/\sigma (\psi)\simeq  0.2 $ at DESY-HERA energies
 at $Q^2=0$ and the ratio is a $Q^2$-dependent quantity as the electroproduction cross sections 
are considered \cite{H1psi2}. 
 The combination of the  energy dependence of the dipole cross section, and the node of the
 radial wavefunction of $2S$ states lead to an anomalous $Q^2$ and energy dependence
 of diffractive production of $2S$ vector mesons \cite{Nemchik:1996cw}. 
Such anomaly appears also
in the $t$-dependence of the differential cross section of radially excited $2S$ light 
vector mesons \cite{Nemchik:2000dd}, which is in contradiction with the usual monotonical 
behavior of the corresponding $1S$ states.
\\
This paper, focuses on the photoproduction of radially excited vector mesons off nuclei in heavy 
ion relativistic collisions, analysing the exclusive photoproduction
 of $\psi^{\prime}$  off nuclei, $\gamma A \rightarrow \psi(2S) X$, where for the coherent scattering
 one has $X=A$, whereas for the incoherent case $X=A^*$ with $A^*$ being an excited state of the
 $A$-nucleon system.
The light-cone dipole formalism is considered. In such framework, the $c\bar{c}$ fluctuation 
of the incoming quasi-real photon
interacts with the nucleus target through the dipole cross section and the result is projected on the
wavefunction of the observed hadron. 
In the high energy regime, the dipole cross
section depends on the gluon distribution in the target and nuclear shadowing of the gluon
distribution is expected to reduce it compared to a proton target.

The ALICE Collaboration
opens the possibility to investigate
 small-$x$ physics with heavy nuclei through measures
 of the diffractive $\psi(1S)$ vector meson 
production at a relatively large rapidity $y\simeq 3$ \cite{ALICE1}  and central rapidities
 \cite{ALICE2} as well in the $\sqrt{s}=2.76$ TeV run.
The incoherent $\psi (1S)$ cross section 
has been also measured \cite{ALICE2}. For nuclear targets, the saturation is enhanced
 i.e.  $Q_{\mathrm{sat}}\propto A^{1/3}$. The LHCb Collaboration has also
 measured the cross section in proton-proton collisions at $\sqrt{s}=7$ TeV  of exclusive dimuon
 final states, including the $\psi (2S)$ state \cite{LHCb}. The ratio at forward rapidity
 $2.0\leq \eta_{\mu^{\pm}}\leq 4.5$ in that case is
 $\sigma(\psi(2S))/\sigma(\psi(1S))= 0.19\pm 0.04$, which is still consistent to the color dipole
 approach formalism.
Therefore it is interesting to investigate the photoproduction of $\psi(2S)$ in
PbPb collisions at the LHC.


\section{Photon-pomeron process in relativistic AA collisions}
\label{coerente}

\hspace{0.5cm}At large impact parameter and at ultra relativistic energies
 nucleus-nucleus collisions are dominated by electromagnetic interaction.
In  heavy ion colliders, the heavy nuclei give rise
 to strong electromagnetic fields due to the coherent action of all protons in the nucleus, which
 can interact with each other. 
Consequently, the total cross section for a given process can be
 factorized in terms of the equivalent flux of photons of the hadron projectile and the 
photon-photon or photon-target production cross section \cite{upcs}.
For the photoproduction of radially excited vector mesons, photon-hadron processes are relevant.
Considering that the photoproduction
is not accompanied by hadronic interaction an analytic expression for the equivalent photon flux
 of a nuclei can be calculated \cite{upcs}
 $\frac{dN_{\gamma}\,(\omega)}{d\omega}  =  \frac{2\,Z^2\alpha_{em}}{\pi\,\omega}\, \left[\xi_R^{AA}\,K_0\,(\xi_R^{AA})\, K_1\,(\xi_R^{AA}) \right.
 -  \left. \frac{(\xi_R^{AA})^2}{2}\,K_1^2\,(\xi_R^{AA})-  K_0^2\,(\xi_R^{AA}) \right].$
where  $\omega$ is the photon energy,  $\gamma_L$ is the Lorentz boost  of a single beam and $K_0(\xi)$ and  $K_1(\xi)$ are the
modified Bessel functions.
Considering symmetric nuclei having radius $R_A$, one has $ \xi_R^{AA}=2R_A\omega/\gamma_L$. 


Using the relation with the photon
energy $\omega$, i.e. $y\propto \ln \, (2 \omega/m_X)$,
the rapidity distribution $y$ for quarkonium photoproduction in nucleus-nucleus collisions can be 
also computed. Explicitly, the rapidity distribution is written down as,
$\frac{d\sigma \,\left[A A \rightarrow   A\otimes \psi(2S) \otimes X \right]}{dy} = \omega \frac{dN_{\gamma} (\omega )}{d\omega }\,\sigma_{\gamma A \rightarrow \psi(2S) X }\left(\omega \right),$
\label{dsigdy}
where $\otimes$ represents the presence of a rapidity gap.
Consequently, given the photon flux, the rapidity distribution is thus a direct measure of the 
photoproduction cross section for a given energy.

In the light-cone dipole frame most of the energy is
carried by the hadron, while the photon has
just enough energy to dissociate into a quark-antiquark pair
before the scattering.
In this representation the probing
projectile fluctuates into a
quark-antiquark pair (a dipole) with transverse separation
$r$ long after the interaction, which then
scatters off the hadron \cite{nik}.
In this picture the amplitude for vector meson production off nucleons reads as
(See e.g. Refs. \cite{nik,mesons})
$\, {\cal A}\, (x,Q^2,\Delta)  = \sum_{h, \bar{h}}
\int dz\, d^2 r \,\Psi^\gamma_{h, \bar{h}}\,{\cal{A}}_{q\bar{q}} \, \Psi^{V*}_{h, \bar{h}} \, ,$
where $\Psi^{\gamma}_{h, \bar{h}}(z,\,r,Q^2)$ and $\Psi^{V}_{h,
  \bar{h}}(z,\,r)$ are the light-cone wavefunctions  of the photon  and of the  vector meson,
 respectively, 
$h$ and $\bar{h}$ are the quark and antiquark helicities,  
$r$ defines the 
relative transverse separation of the pair (dipole), $z$ $(1-z)$ is the
longitudinal momentum fractions of the quark (antiquark), $\Delta$ denotes 
the transverse momentum lost by the outgoing proton ($t = - \Delta^2$) and $x$ is the Bjorken
variable.

The predictions presented here take into account the corrections due to skewedness
effect (off-diagonal gluon exchange) and real part of amplitude.
Detail on the model dependence on these corrections can be found for instance in Ref. \cite{GM}.

The photon wavefunctions are relatively well known \cite{mesons}.
For the meson wave function the boosted gaussian
 wavefunction was considered \cite{Sandapenpsi}.

The boosted gaussian wavefunction
 considered here is a simplification of the NNPZ wavefunction presented in
 Refs. \cite{nik,Nemchik:1996pp}. 
It has been compared to recent analysis of DESY-HERA data 
for vector meson exclusive processes \cite{Sandapenpsi},\cite{SandaUps}. 
The node effect plays an important role in the description of the
 measured ratio $\sigma (\psi^{\prime})/\sigma (J/\psi)$ in the photoproduction case.
 Such a ratio is sensitive to the time-scale of the production process. In the dipole approach
 the interactions occur during the period where the color dipole is compact having a transverse
 size $r \simeq 1/m_q$ and the production cross section is proportional to the square of the
 quarkonium wavefunction at origin, $\sigma \propto |\phi (0)|^2$. On the other hand, further
 interactions depend on the wavefunction profile for transverse sizes larger than
 $r_B = {\cal O}(1/\alpha_s m_q)$, the Bohr radius.
 In exclusive charmonia
 electroproduction at relatively large $Q^2$ the dipole size is of order $1/Q^2\ll r_B$ and
 the cross section is predicted to be proportional to $|\phi_n (0)|^2$. This leads the ratio
 to be of order $|\phi_{2S} (0)|^2/|\phi_{1S} (0)|^2\simeq 0.6$ at large $Q^2$ whereas the
 measured value in photoproduction is around 0.16 \cite{H1psi2}. 
 The moderate value of charm mass and
 the dominant color transparency behavior of dipole cross section $\sigma_{dip}\propto r^2$
 imply the amplitude to probe the meson wavefunction at a transverse size around $r_B$ \cite{nik,Nemchik:1996pp}.
So, the node effect reduces the $\psi (2S)$ contribution and the DESY-HERA ratio is correctly described.
At $Q^2\rightarrow 0$ the leading logarithmic
approximation $rl_{\perp} \ll 1$, which gives the usual $\sigma_{dip}\propto r^2$, is not 
able to provide alone the correct value for the ratio $\psi^{\prime}/\psi$ \cite{Suzuki}.
Here, $l_{\perp}$ is the exchanged gluon transverse momentum in a two-gluon exchange model. 
Therefore, important contributions come from the overlap of the large-sized color dipole
 configurations and the $\psi (2S)$ wavefunction. Thus, despite the leading logarithmic
 approximation to be able to describe the $J/\psi$ production cross section the same is not true
 for the excited states as the  $\psi^{\prime}$.
 For this reason it was used a dipole cross section model that takes into account the correct 
behavior for large dipole configurations (the transition hard-soft is given by the saturation scale). 

The exclusive $\psi(2S)$ photoproduction off nuclei for coherent and incoherent processes can be 
simply computed in high energies where the large coherence length $l_c\gg R_A$ is fairly valid.
 In such case the transverse size of $c\bar{c}$ dipole is frozen by Lorentz effects.
 The expressions for the coherent and incoherent cross sections are given by \cite{Boris},
\begin{eqnarray}
\sigma_{coh}^{\gamma A} & = & \int d^2b\, |\langle \Psi^V|1-\exp\left[-\frac{1}{2}\sigma_{dip}(x,r) T_A(b)  \right]|\Psi^{\gamma}\rangle |^2, \label{eq:coher}
 \\
\sigma_{inc}^{\gamma A} & = & \frac{1}{16\pi\,B_V(s)}\int d^2b\,T_A(b) 
  \times  |\langle \Psi^V|\sigma_{dip}(x,r) \exp\left[-\frac{1}{2}\sigma_{dip}(x,r)T_A(b)  \right]|\Psi^{\gamma}\rangle|^2. 
\label{eq:incoh}
\end{eqnarray} 
where $T_A(b)= \int dz\rho_A(b,z)$  is the nuclear thickness function given by integration of
 nuclear density along the trajectory at a given impact parameter $b$. In addition, $B_V$ is
 the diffractive slope parameter in the reaction $\gamma^*p\rightarrow \psi p$. Here, we consider
 the energy dependence of the slope using the Regge motivated expression
 $B_V(W_{\gamma p})=b_{el}^V + 2\alpha^{\prime}\log \frac{W_{\gamma p}^2}{W_0^2}$ with
 $\alpha^{\prime}=0.25$ GeV$^{-2}$ and $W_0=95$ GeV.  It is used the measured slopes 
\cite{H1psi2} for $\psi(1S)$ and $\psi(2S)$ at $W_{\gamma p}=90$ GeV, i.e.
 $b_{el}^{\psi(1S)}= 4.99\pm 0.41$ GeV$^{-2}$ and $b_{el}^{\psi(2S)}= 4.31\pm 0.73$ GeV$^{-2}$,
 respectively.

For the dipole cross section was considered the Color Glass Condensate model \cite{IIM}
 for $\sigma_{dip}(x,r)$. This model has been tested for a long period against DIS, diffractive
 DIS and exclusive production processes in $ep$ collisions.
 Corrections due to gluons shadowing were also considered
 as the gluon density in nuclei
 at small-$x$ region is known to be suppressed compared to a free nucleon. 
That is, we will
 take $\sigma_{dip}\rightarrow R_G(x,Q^2,b)\sigma_{dip}$  following studies in Ref. \cite{Borispsi}.
 The factor $R_G$ is the nuclear gluon density ratio. In the present investigation we will use
 the nuclear ratio from the leading twist theory of nuclear shadowing based on generalization
 of the Gribov-Glauber multiple scattering formalism as investigated in Ref. \cite{GFM}.
 We used the two models available for $R_G(x,Q^2)$ in \cite{GFM}, Models 1 and 2, which 
correspond to higher nuclear shadowing and lower nuclear shadowing, respectively.

\section{Results and discussions}
\label{resultados}

\begin{figure}
\centering
\includegraphics[scale=0.35]{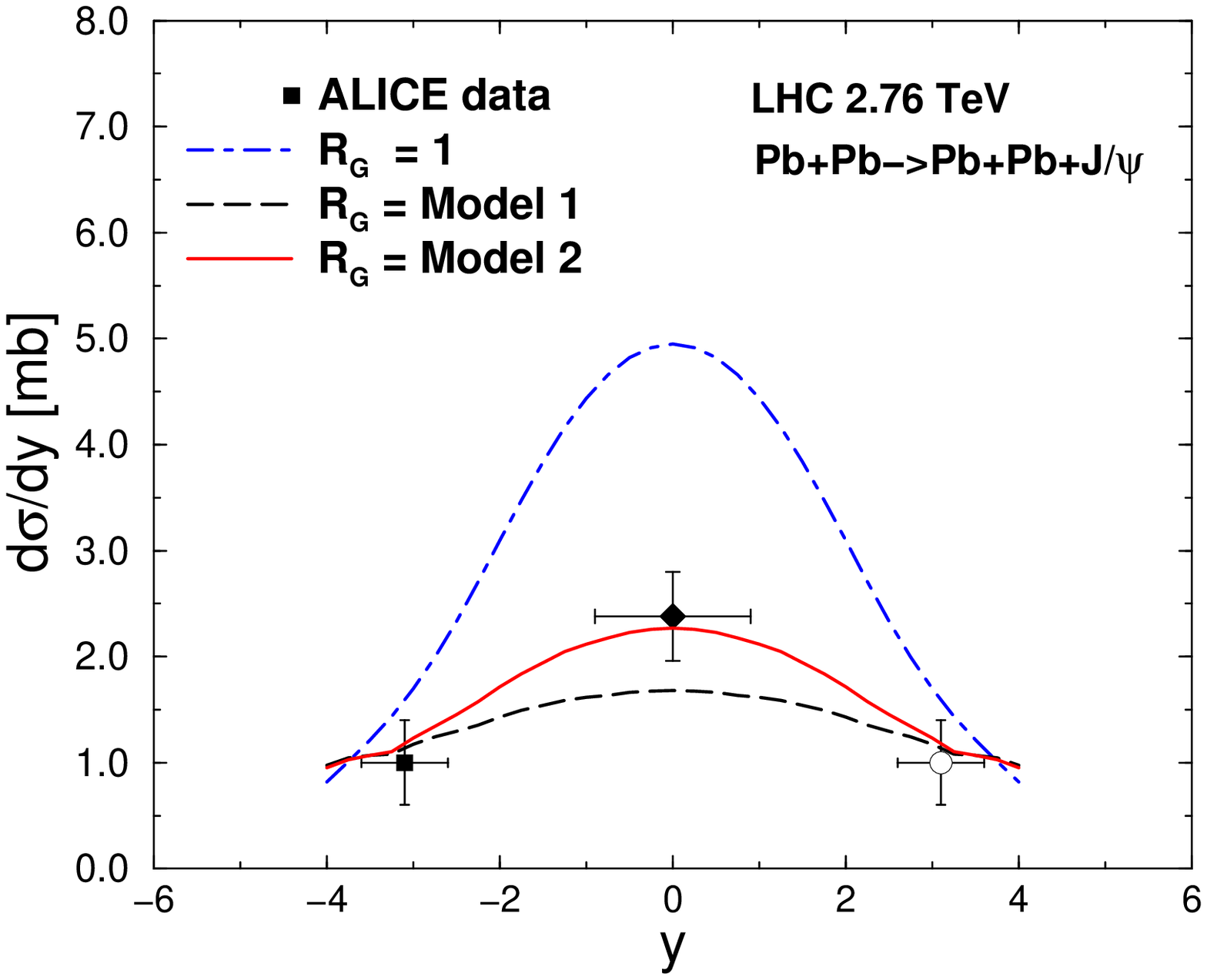} 
\includegraphics[scale=0.35]{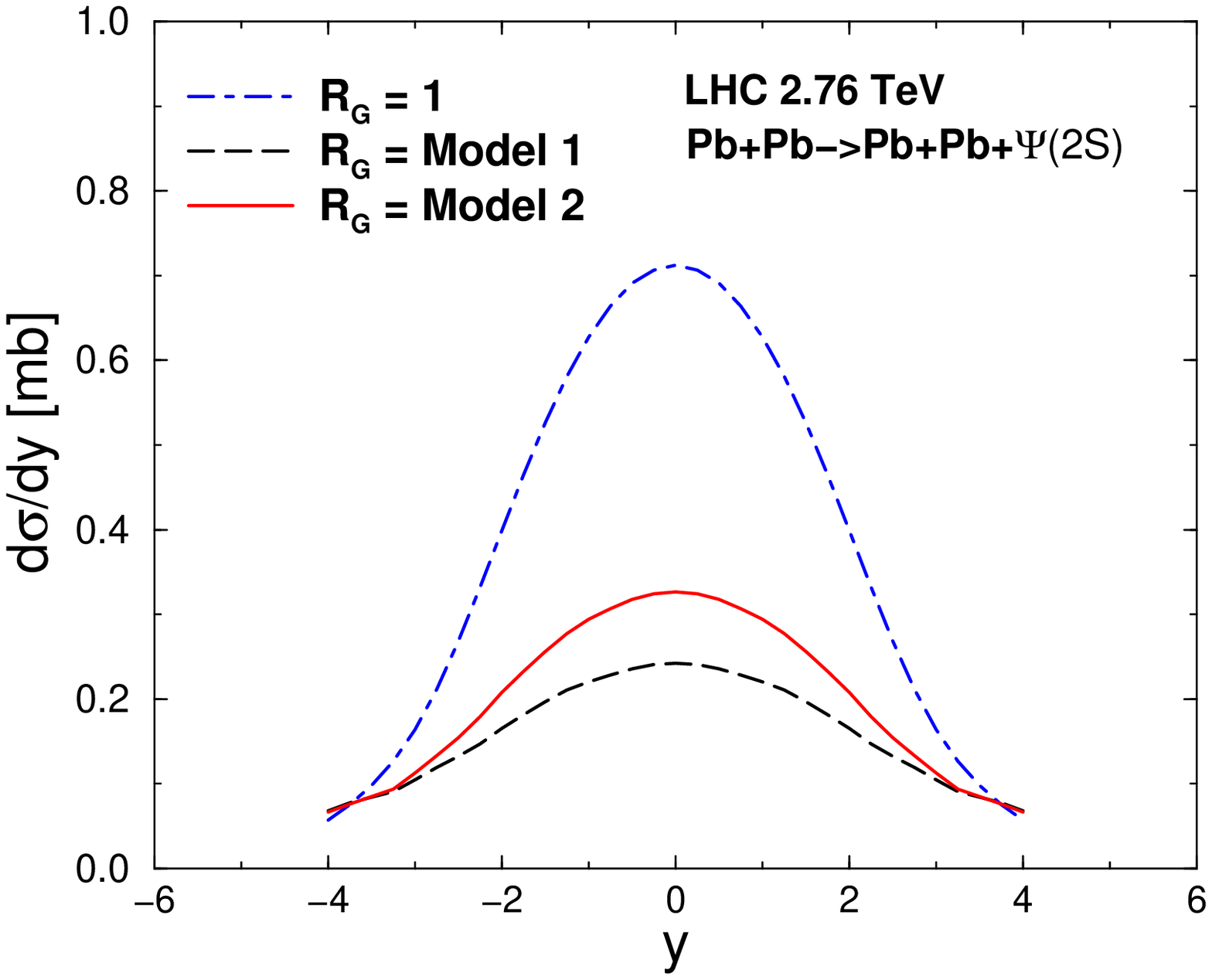}
\caption{(Color online) \it The rapidity distribution of coherent $\psi(1S)$ meson photoproduction at $\sqrt{s}=2.76$ TeV in PbPb collisions at the LHC \cite{nos}. The theoretical curves stand for color dipole formalism using $R_G=1$ (dot-dashed curve) and two scenarios for the nuclear gluon distribution (solid and long-dashed curves, see text). Data from ALICE collaboration \cite{ALICE1,ALICE2}.}
\label{psi1s}
\end{figure}

\hspace{0.5cm}The Fig. 1 (left) presents the numerical calculations for the rapidity
 distribution of coherent $\psi(1S)$ state within the color dipole formalism, Eqs. (\ref{dsigdy}) 
and (\ref{eq:coher}), using distinct scenarios for the nuclear gluon shadowing \cite{nos}. The dot-dashed curve
 represents the result using $R_G=1$ and it is consistent with previous calculations using the same
 formalism \cite{GM}. 
The ALICE data is overestimated on the backward (forward) and mainly in central
 rapidities. 
The threshold factor for $x\rightarrow 1$ was not included in the present calculation, so
the overestimation in the backward/forward rapidity case, is already expected.
 In that kinematical region either a small-$x$ photon scatters off a large-$x$ gluon or vice-versa.
 For instance, for $y\simeq \pm 3$ one gets $x$ large as $0.02$. On the other hand, for central rapidity 
$y=0$ one can be obtained $x=M_Ve^{\pm y}/\sqrt{s_{\mathrm{NN}}}$ smaller than $10^{-3}$ for the 
nuclear gluon distribution.
The ALICE data \cite{ALICE2} is 
overestimated by a factor 2 considering $R_G=1$, as already noticed in the recent study of Ref. \cite{LM}. 
If we consider nuclear shadowing renormalizing the dipole cross section, the situation is improved 
due to the gluon density in nuclei at small Bjorken $x$ is expected to be suppressed compared to a free
 nucleon due to interferences. 
For the ratio of the gluon density, $R_G(x,Q^2=m_V^2/4)$, we have
 considered the theoretical evaluation of Ref. \cite{GFM}. There, two scenarios for the gluon 
shadowing are investigated: Model 1 corresponds to a strong gluon shadowing and Model 2 concerns to 
small nuclear shadowing. The consequence of renormalizing the dipole cross section by gluon shadowing
 effects is represented  by the long-dashed (Model 1) and solid (Model 2) lines, respectively. 
In the current analysis, the small shadowing option is preferred . The theoretical uncertainty related to the choice of meson wavefunction is relatively large.
 As a prediction at central rapidity, one obtains $\frac{d\sigma}{dy} (y=0) = 4.95, \,1.68$ and 
$2.27$ mb for calculation using $R_G=1$, Model 1 and Model 2, respectively.
$R_G$ was considered as independent on the impact parameter. It is known long time ago 
that a $b$-dependent ratio could give a smaller suppression compared to our calculation.
 For instance, in Ref. \cite{Borispsi} the suppression is of order 0.85 for the LHC energy and
 central rapidity. 

The Fig. 2 (right) shows the first estimate in literature for the coherent photoproduction
 of $\psi(2S)$ state in nucleus-nucleus collisions \cite{nos}. 
  The theoretical predictions follow the general trend as for the $1S$ state, where the
 notation for the curves are the same as used in Fig. 1 (left). In particular, for $R_G=1$ one
 obtains for central rapidity $\frac{d\sigma}{dy}(y=0) = 0.71$ mb and the following in the
 forward/backward region  $\frac{d\sigma}{dy}(y=\pm 3) = 0.16$ mb. When introducing the suppression 
in the dipole cross section due nuclear shadowing one gets instead $\frac{d\sigma}{dy}(y=0) = 0.24$ mb
 and  0.33 mb  for Model 1 and Model 2, respectively. At central rapidities, the meson state ratio
 is evaluated to be
 $R_{\psi}^{y=0}=\frac{\sigma_{\psi(2S)}}{dy}/\frac{d\sigma_{\psi(1S)}}{dy} (y=0)= 0.14$ in case 
$R_G=1$ which is consistent with the ratio measured in CDF, i.e. $0.14\pm 0.05$, on the
 observation of exclusive charmonium production at 1.96 TeV in $p\bar{p}$ collisions \cite{CDF}.
 A similar ratio is obtained using Model 1 and Model 2 at central rapidity as well. As a prediction
 for the planned LHC run in PbPb mode at 5.5 TeV, we obtain
 $\frac{d\sigma_{coh}}{dy}(y=0) = 1.27$ mb and $\frac{d\sigma_{inc}}{dy}(y=0) = 0.27$ mb  for the
 coherent and incoherent $\psi(2S)$ cross sections (upper bound using $R_G=1$), respectively.


The Fig. 3 presents the incoherent contribution to the rapidity distribution for
 both $\psi(1S)$ (solid line) and $\psi (2S)$ (dashed line) meson states \cite{nos}.
 For the $\psi(1S)$ state, the present
 calculation can be directly compared with those studies presented in Ref. \cite{LM}.
The incoherent cross section  $\frac{d\sigma_{\mathrm{inc}}}{dy}$  ranges between
 0.5 to 0.7 mb (using IIM dipole cross section) or between 0.7 to 0.9 mb (using fIPsat dipole cross
 section) at central rapidities, with the uncertainty determined by the distinct meson wavefunction
 considered \cite{LM}.
 Here, was obtained  $\frac{d\sigma_{\mathrm{inc}}}{dy}(y=0) = 1.1$ mb using a
 different expression for the incoherent amplitude, Eq. (\ref{eq:incoh}). 
This result describes
 the recent ALICE data \cite{ALICE2} for the incoherent cross section at mid-rapidity,
 $\frac{d \sigma_{inc}^{\mathrm{ALICE}}}{dy} (-0.9<y<0.9) = 0.98 \pm 0.25$ mb. 
For the $\psi(2S)$ state, was found $\frac{d\sigma_{\mathrm{inc}}}{dy} = 0.16$ mb for central
 rapidities. In both cases was only computed the case for $R_G=1$. Therefore, this gives an
 upper bound for the incoherent cross section compared to Model 1 and Model 2 calculation.
 For the incoherent case, the gluon shadowing is weaker than the coherent case
 and the reduction is around 20 \% compared to the case $R_G=1$. The incoherent
  piece is quite smaller compared to the main coherent contribution. As an example of order of
 magnitude, the ratio incoherent/coherent is a factor 0.22 for the $1S$ state and 0.23 for the
 $2S$ state at central rapidity. 

\begin{figure}
\begin{center}
\includegraphics[scale=0.35]{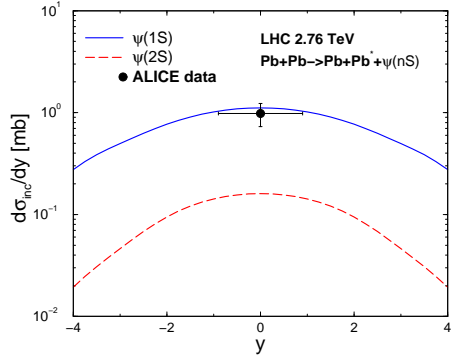}
\caption{(Color online) \it The rapidity distribution of incoherent $\psi(1S)$ (solid line) and $\psi(2S)$ (dashed line) meson photoproduction at $\sqrt{s}=2.76$ TeV in PbPb collisions at the LHC \cite{nos}. Data from ALICE collaboration \cite{ALICE2}.}
\end{center}
\label{fig:3}
\end{figure}

\section{Conclusions}
\label{conc}

\hspace{0.5cm}The photoproduction of radially excited vector mesons was investigated in heavy ion
 relativistic collisions as the $\psi(2S)$ charmonium state using the light-cone dipole formalism.
 Predictions are done for PbPb collisions at the CERN-LHC energy of 2.76 TeV.
The fact that the gluons shadowing suppresses
 the dipole cross section was studied and the results for $R_G=1$ gives the larger
 cross sections. The coherent exclusive photoproduction of $\psi(2S)$  off nuclei
 has an upper bound of order 0.71 mb at $y=0$ down to 0.10 mb for backward/forward rapidities
 $y=\pm 3$. The incoherent contribution was also computed and it is a factor 0.2 below the coherent
 one.  
A small nuclear shadowing
 $R_G(x,Q^2=\frac{m_V^2}{4})$ is preferred in ALICE data description whereas the usual $R_G=1$ value 
overestimates the central rapidity cross section by a factor 2 for the
$\psi(1S)$ state photoproduction.
For incoherent cross section, the present
 theoretical approach describes the ALICE data. Thus, 
the central rapidity data measured by ALICE Collaboration for the rapidity distribution of 
the $\psi(1S)$ state is crucial to constrain the nuclear gluon function.
 The cross section for 
exclusive quarkonium  production is proportional to $[\alpha(Q^2)xg_A(x,Q^2)]^2$ in the leading-order
 pQCD calculations, evaluated at the relevant scale $Q^2\approx m_V^2/4$ and at momentum
 fraction $x\simeq 10^{-3}$ in central rapidities. The theoretical uncertainty is large and it
 has been investigated in several studies \cite{AGG,AB}. Along these line, the authors of
 Ref. \cite{GKSZ} extract the nuclear suppression factor, $S(x\approx 10^{-3})=0.61 \pm 0.064$,
 using the ALICE data on coherent $\psi (1S)$ and considering the nuclear gluon shadowing predicted
 by nuclear pdf's and by leading twist nuclear shadowing.

\end{document}

%% file: econfmacros.tex
\def\beq{\begin{equation}}
\def\eeq#1{\label{#1}\end{equation}}
\def\eeqn{\end{equation}}

\def\beqa{\begin{eqnarray}}
\def\eeqa#1{\label{#1}\end{eqnarray}}
\def\eeqan{\end{eqnarray}}

\let\bar=\overbar

\def\Dslash{\not{\hbox{\kern-4pt $D$}}}
\def\dslash{\not{\hbox{\kern-2pt $\del$}}}

\def\msb{{\bar{\ssstyle M \kern -1pt S}}}

